\documentclass[preprint2]{aastex}
\usepackage{rotating}
\usepackage{sidecap}
\usepackage{longtable}
\usepackage{amsmath} 
\usepackage{url}
\pdfoutput=1 

\newcommand{\mjup}{M$_{\text{J}}$}

\begin{document}

\title{Mapping the shores of the brown dwarf desert. IV. Ophiuchus}

\author{Anthony C. Cheetham\altaffilmark{1,2,*}, Adam L. Kraus\altaffilmark{3}, Michael J. Ireland\altaffilmark{4}, Lucas Cieza\altaffilmark{5,6}, Aaron C. Rizzuto\altaffilmark{3}, Peter G. Tuthill\altaffilmark{2}}
\altaffiltext{*}{anthony.cheetham@unige.ch}
\altaffiltext{1}{Observatoire de Gen\`{e}ve, Universit\'{e} de Gen\`{e}ve, 51 chemin des Maillettes, 1290, Versoix, Switzerland}
\altaffiltext{2}{Sydney Institute for Astronomy, School of Physics, University of Sydney, NSW 2006, Australia}
\altaffiltext{3}{Department of Astronomy, The University of Texas at Austin, Austin, TX 78712, USA}
\altaffiltext{4}{Research School of Astronomy \& Astrophysics, Australian National University, Canberra ACT 2611, Australia}
\altaffiltext{5}{Nucleo de Astronomia, Universidad Diego Portales, Av. Ej\'{e}rcito 441, Santiago, Chile}
\altaffiltext{6}{Millennium Nucleus Protoplanetary Disks in ALMA Early Science, Universidad Diego Portales, Av. Ejercito 441, Santiago, Chile}

\begin{abstract}
We conduct a multiplicity survey of members of the $\rho$ Ophiuchus cloud complex with high resolution imaging to characterize the multiple star population of this nearby star forming region and investigate the relation between stellar multiplicity and star and planet formation. Our aperture masking survey reveals the presence of 5 new stellar companions beyond the reach of previous studies, but does not result in the detection of any new substellar companions. We find that 43$\pm6$\% of the 114 stars in our survey have stellar mass companions between 1.3-780\,AU, while 7$^{+8}_{-5}$\% host brown dwarf companions in the same interval. By combining this information with knowledge of disk-hosting stars, we show that the presence of a close binary companion (separation $<40$\,AU) significantly influences the lifetime of protoplanetary disks, a phenomenon previously seen in older star forming regions. At the $\sim$1-2\,Myr age of our Ophiuchus members $\sim$2/3 of close binary systems have lost their disks, compared to only $\sim$30\% of single stars and wide binaries. This has significant impact on the formation of giant planets, which are expected to require much longer than 1\,Myr to form via core accretion and thus planets formed via this pathway should be rare in close binary systems.
\end{abstract}



\section{Introduction}

Multiplicity surveys provide some of the most stringest tests of stellar formation theories. The distribution and frequency of companions are key predictions of these theories that are easily tested. Radial velocity (RV) surveys have had great success at revealing close stellar and substellar companions, especially around older stars, while high resolution imaging surveys have increased our understanding of companions at wider separations.

RV surveys have revealed a wealth of both stellar and planetary mass companions in close orbits around their host stars. However, they have revealed a surprising lack of brown dwarf mass companions in close orbits, with estimates suggesting frequencies of $<$ 1\% \citep{2000PASP..112..137M,2006ApJ...640.1051G}. This phenomenon has been labelled the ``brown dwarf desert''.

In contrast, imaging surveys targeting wider separations have found that the frequency of such companions may not be anomalously low, but rather an extension of the binary mass-ratio function to lower masses \citep{2009ApJS..181...62M}.

In this paper we investigate the crucial separations between these two approaches, following on from previous work in \cite{kraus08} and \cite{2011ApJ...731....8K}. These studies investigated the Upper Scorpius subgroup of the Sco-Cen OB association and the Taurus-Auriga star forming region respectively. Both studies resulted in companion distributions consistent with a flat mass ratio distribution, finding 5 and 6 companions with mass ratios q$\le0.1$ respectively. This provides further evidence that the frequency of brown dwarf companions is consistent with the observed trend for stellar mass companions.

The distributions of orbital parameters in binary stars offers another key test of star formation models, through comparison of their predictions to the observed properties of multiple systems. Several studies have targeted binaries in the field, finding evidence that these properties are mass-dependent \citep{1991A&A...248..485D,1992ApJ...396..178F,2003ApJ...587..407C,2010ApJS..190....1R}. For example, companions to Solar mass stars have a higher mean separation than companions to low mass stars. In addition, the number of low mass companions to Solar mass stars appears high, while the distribution of companion mass ratios for low mass stars is peaked more towards unity.

However, surveys of young star forming regions have shown significant differences in many of these properties. The overall binary frequencies in these regions appear to be much higher than the field \citep[$\gtrsim 80$\%][]{1993AJ....106.2005G,1995ApJ...443..625S,2000A&A...356..541K,kraus08,2011ApJ...731....8K}, and the shapes of the separation and mass ratio distributions appear different. Dynamical interactions may play a significant role in causing these differences, since studies of denser regions such as young clusters have shown similar results to the field \citep{1999A&A...343..831D,2006A&A...458..461K}. A large fraction of stars in the field are thought to originate from these dense clusters, and the similarities between their properties echoes this idea.

Samples less affected by dynamical interactions provide simpler tests of binary formation processes, and so obtaining robust statistics for young star forming regions like Ophiuchus are important to test multiple star formation theories.

The relationship between binarity and disk evolution is another key link to understanding star and planet formation. While a large body of work has concentrated on the evolution and formation of planets in single star systems, the majority of solar type stars exist in multiple systems which may have a profound effect on the way in which these processes occur.

Surveys targeting disks and stellar multiplicity in nearby star forming regions have shown correlations between the presence of a binary companion and the properties and presence of a circumstellar disk \citep{1997ApJ...490..353G,2009ApJ...696L..84C,2010ApJ...709L.114D,2012ApJ...745...19K}. These studies found that the presence of a close ($\le 40$\,AU) binary companion can significantly speed up the dispersal or inhibit the formation of protoplanetary disks.

By combining the results of previous multiplicity surveys of several star forming regions, \cite{2009ApJ...696L..84C} found that close binaries with separations less than 40\,AU were half as likely to retain their disks as binary systems with larger separations. However, the timescale of this effect remains unclear.

Comparing the results for different star forming regions by age, \cite{2012ApJ...745...19K} found that $\sim$2/3 of all close binaries have no disk at ages of 1-2\,Myr in Taurus. Despite this, stable configurations appear to exist that allow some disks around close binaries to persist for $\sim10$\,Myr in Upper Scorpius. In contrast, the presence of a wider companion does not appear to affect the lifetime of protoplanetary disks at these ages. Comparison of results from the 1-2\,Myr old Ophiuchus region with the regions investigated in previous studies will provide information on the age dependent properties of these effects and the scatter between individual regions.

The relationship between the timescales of disk dispersal and giant planet formation defines the relative abundance of such planets. The two canonical giant planet formation theories of core accretion \citep{1996Icar..124...62P} and disk instability \citep{2001ApJ...563..367B} predict different timescales. Core accretion requires several Myr to form giant planets from protoplanetary disks \citep{2005Icar..179..415H}, while disk instability is most efficient at much younger ages ($\lesssim$0.5\,Myr, \citealt{2001ApJ...563..367B}). The short lifetime of disks in close binary systems would lead to a low occurence rate of giant planets that formed through core accretion, a key prediction to differentiate between the two theories.

We utilise the technique of Sparse Aperture Masking (SAM) to perform a high-resolution survey of the nearby Ophiuchus star forming region. The relatively small distance to this association as well as its relative youth provide an ideal opportunity to investigate changes in both the prevalence of brown dwarf companions and the relationship between disk evolution and multiplicity. Comparison of results from the $\sim$1-2\,Myr old Ophiuchus region with the older regions investigated in previous studies will provide information on the age dependent properties of these effects.


\section{Survey Sample} \label{sec:targs}

\begin{figure*}
	\begin{center}
	\includegraphics[width=0.9\textwidth]{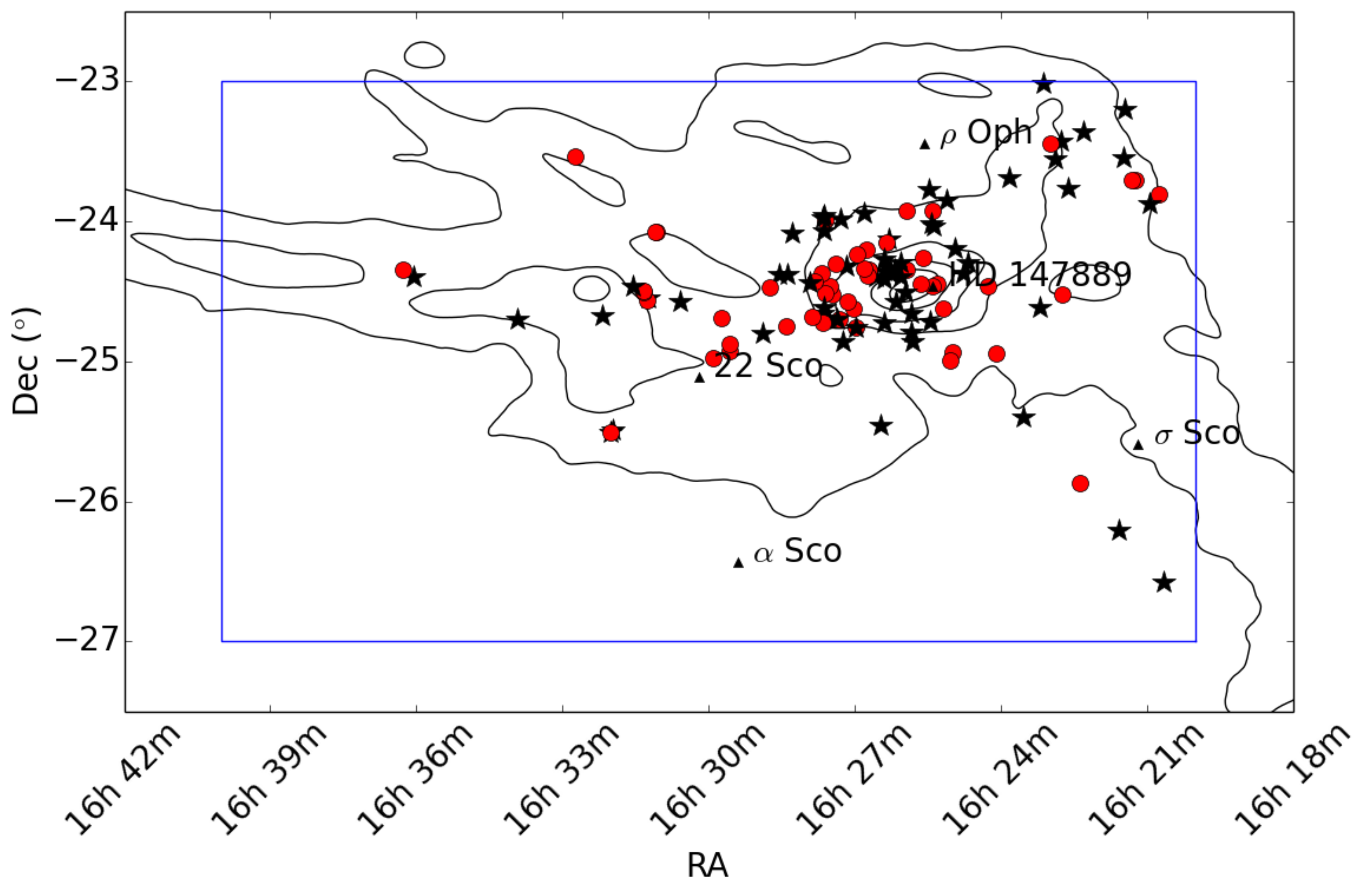}
	\caption{Distribution of targets that fit the criteria described in Section \ref{sec:targs}, plotted over contours marking 1, 2.5, 5, 10 and 15\,mag of (B-V) extinction from \cite{1998ApJ...500..525S}. Several prominent nearby stars are marked with triangles and labelled for reference, as well as HD 147889, the highest mass target in our sample. Targets marked with red circles were not obseved with aperture masking.}
	\label{fig:dust_map}
	\end{center}
\end{figure*}

\begin{figure}[h]
	\begin{center}
	\includegraphics[width=0.45\textwidth]{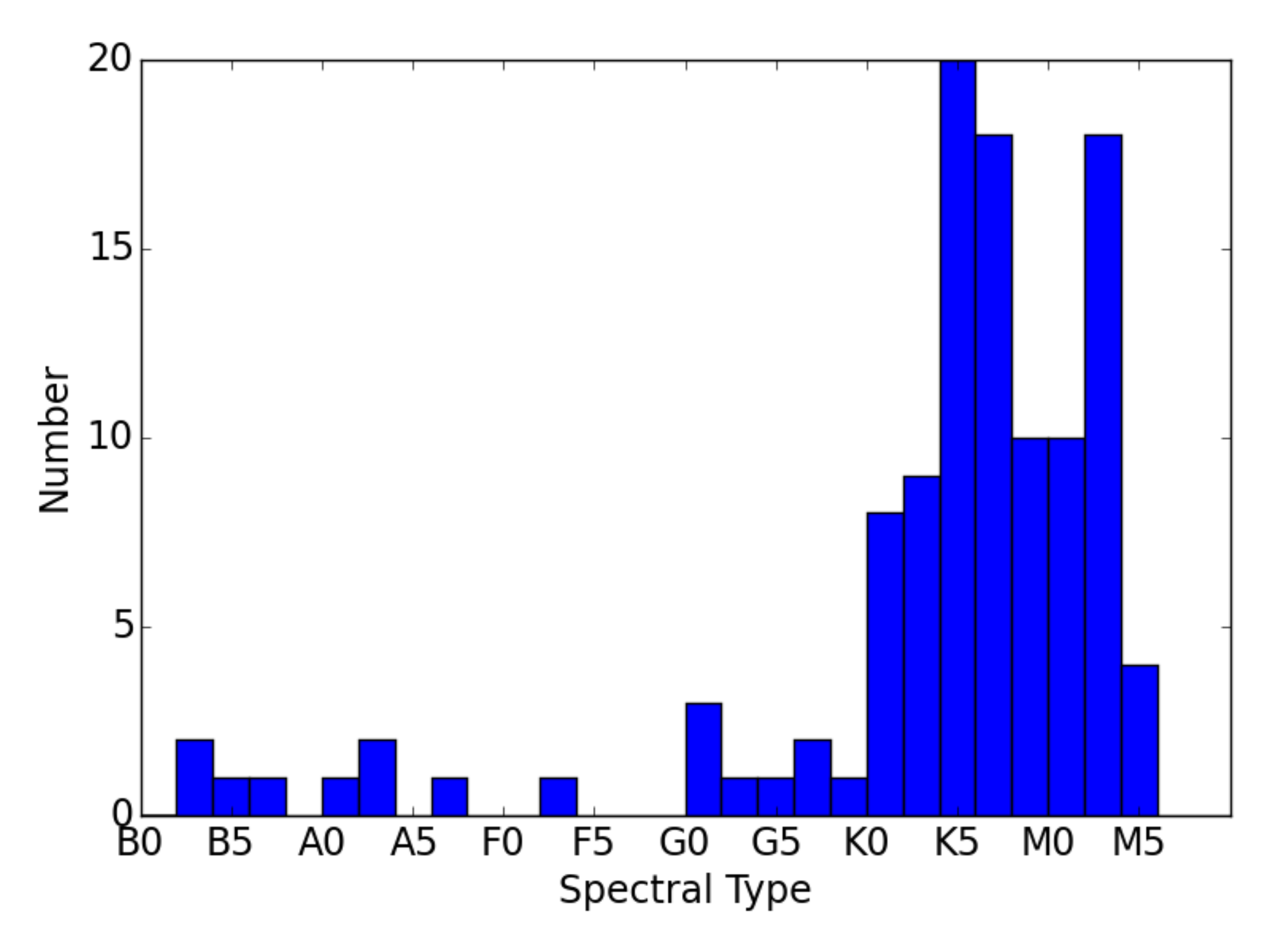}
	\caption{The distribution of spectral types for the input sample described in Section \ref{sec:targs}.}
	\label{fig:spt_hist}
	\end{center}
\end{figure}

This survey consists of observations of members of the $\rho$ Ophiuchus cloud complex. The star formation history of this region is complex, with several periods of recent star formation. Its similar location and distance to the Sco-Cen OB association has led to the suggestion that the most recent episode of star formation in this region was triggered by an interaction between the large L1688 cloud and a shock wave from the Upper Scorpius sub group, occuring approximately 1-1.5\,Myr ago \citep{1977AJ.....82..198V,1986ApJ...306..142L}.

The first estimates of the distance to the cloud complex suggested a value of around 160\,pc \citep{1958ApJ...128..533B,1974MNRAS.168..371W}. However, more recent studies have measured the distance to be between 119-135\,pc \citep{1989A&A...216...44D,2008A&A...480..785L,2008AN....329...10M}. In this study, we adopt a value of 130\,pc for all of our association members.

\begin{deluxetable}{lccccccccc}
\tabletypesize{\scriptsize}
\tablecaption{Survey Sample\label{table:targets}}
\tablehead{\colhead{Name} & \colhead{RA} & \colhead{Dec} & \colhead{SpT} & \colhead{Mass} & \colhead{R} & \colhead{K} & \colhead{Reference}\tablenotemark{a} & {Disk?} & {Multiple} \\ 
\colhead{} & \colhead{(J2000)} & \colhead{(J2000)} & \colhead{} & \colhead{($M_{\odot}$)} & \colhead{(mag)} & \colhead{(mag)} & \colhead{} & \colhead{} & \colhead{System?}}
\startdata
{DoAr 13} & 16 20 39.60 & -26 34 28.4 & M2 & 0.61 & 14.89 & 9.921 & 6 & Y & N \\ 
{GSC 6794-480} & 16 20 45.96 & -23 48 20.8 & K3  & 2.07 & 12.66 & 8.927 & 7 & N & N \\ 
{[MMG98] RX J1620.9-2352} & 16 20 57.87 & -23 52 34.3 & K3 & 2.07 & 9.88 & 8.393 & 6 & N & N \\ 
{[MMG98] RX J1621.2-2342a} & 16 21 14.50 & -23 42 20.0 & K7 & 0.98 & 12.51 & 8.99 & 6 & N & N \\ 
{HIP 80126} & 16 21 19.18 & -23 42 28.7 & B5  & 5.90 & 6.98 & 6.418 & 12 & Y & N \\ 
{[MMG98] RX J1621.4-2312} & 16 21 28.44 & -23 12 11.0 & K7 & 0.98 & 13.55 & 8.739 & 6 & N & N \\ 
{[MMG98] RX J1621.4-2332b} & 16 21 28.81 & -23 32 38.9 & M0 & 0.82 & 11.02 & 7.167 & 6 & N & Y \\ 
{WSB 9} & 16 21 34.69 & -26 12 26.9 & K5 & 1.20 & 12.88 & 8.864 & 5 & Y & N \\ 
{WSB 12} & 16 22 18.52 & -23 21 48.0 & K5 & 1.20 & 13.03 & 8.109 & 12 & Y & N \\ 
{WSB 13} & 16 22 22.31 & -25 52 19.9 & M4 & 0.29 & 14.56 & 9.438 & 5 & N & N \\ 
\enddata

\tablecomments{R magnitudes taken from NOMAD (\cite{zacharias2005naval}), K magnitudes taken from 2MASS (\cite{skrutskie2006two}). For targets whose most recent classification in the literature provided a range of possible spectral types, the mid-range spectral type was used.\\ Table \ref{table:targets} is published in its entirety in the electronic edition of The Astrophysical Journal. A portion is shown here for guidance regarding its form and content.}

\tablenotetext{a}{Spectral Type References:
(1) \cite{elias78}; (2) \cite{herbig88}; (3) \cite{houk1988michigan}; (4) \cite{bouvier92}; (5) \cite{meyer93}; (6) \cite{martin98}; (7) \cite{preibisch98}; (8) \cite{luhman99}; (9) \cite{doppman03}; (10) \cite{wilking05}; (11) \cite{2006A&A...460..695T}; (12) \cite{cieza10}; (13) \cite{erickson11}.}
\end{deluxetable}

Members of Ophiuchus have a wide range of ages consistent with several episodes of star formation in different regions. A young population of Young Stellar Objects (YSOs) in the centre of the cloud with a median age of 0.3\,Myr was identified by \cite{1995ApJ...450..233G} and \cite{luhman99}, but an older population of YSOs covering a larger area has an age of 2\,Myr, as noted by \cite{wilking05}. These studies estimated the age of observed members through the comparison of H-R diagram positions to model evolutionary tracks, a process that can introduce large systematic effects, as illustrated by the analysis of F-type stars in Sco-Cen by \cite{2012ApJ...746..154P}. In this study we adopt an age of 1-2\,Myr and focus on members close to the L1688 cloud core.

Our target list was arrived at by considering membership of the $\rho$ Ophiuchus cloud complex. Since no canonical survey exists, a conservative approach was taken where all possible members from the literature were considered and then subject to a cut in right ascension and declination (16 20 00 to 16 40 00 and -23 00 00 to -27 00 00 respectively). This was chosen to cover a wide region around the L1688 cloud that hosts ongoing star formation. The initial target list was made by compiling targets from \cite{herbig88}, \cite{bouvier92}, \cite{meyer93}, \cite{martin98}, \cite{preibisch98}, \cite{luhman99}, \cite{wilking05}, \cite{cieza10}, and \cite{erickson11}.

Our targets were further restricted to those members with measured photometry and known spectral types. Considering the limiting magnitudes for reasonable AO correction at Keck and the VLT, targets fainter than R = 15 mags and K = 9.5 mags (for the Keck NIRC2 visible WFS and the VLT NACO IR WFS respectively) were not included. In an attempt to give a sample with better mass completeness, stars with spectral types later than M4 were observed but excluded from statistical analysis.

Out of the 236 possible Ophiuchus members from the literature, 114 fit these criteria. These targets are listed in Table \ref{table:targets}, and their distribution on the sky is shown overlaid on an extinction map of the Ophiuchus L1688 cloud in Figure \ref{fig:dust_map}. The distribution of spectral types for these targets is shown in Figure \ref{fig:spt_hist}. This target list comprises the input sample for our study, and consists of observed targets, known binaries that were not reobserved, and targets that were not observed.

We have compiled the results of previous multiplicity surveys of $\rho$ Ophiuchus members performed by \cite{1995ApJ...443..625S}, \cite{2002AJ....124.1082K}, \cite{2003ApJ...591.1064B}, \cite{2005A&A...437..611R} and \cite{cieza10}. The 0.9\,arcsec binary [MMG98] RX J1622.7-2325a was also added. This companion was found by \cite{2007ApJ...657..338P}, from AO images taken in conjunction with their spectroscopic observations. Both components were spectrally classified, and so we adopt a contrast ratio that preserves this classification. In total, we found 40 companions to members of our target list (8 in triple systems and 32 in binary systems).

Due to difficulties with observing wide equal binary systems the majority of known multiple systems were not reobserved, which will introduce a bias against the detection of higher order systems. The exceptions to this were the stars ROXs 12, ROXs 42B, ROXs 47A, EM* SR 20, EM* SR 21, EM* SR 24 S-N, GSS 31 and [MMG98] RX J1622.7-2325a. The secondaries in these systems were either wide enough to allow the individual components to be studied separately (separation $\gtrsim4"$), close enough to have little effect on the performance of the AO system (separation $\lesssim1"$), or faint enough to allow the AO system to lock on the primary (contrast ratio $\lesssim0.5$). The EM* SR 24 S-N and ROXs 43A-B components were bright enough to make our target list individually and are included separately.

Two of our targets host known wide companions with masses close to or below the deuterium burning limit (ROXs 42B and ROXs 12). Both have been confirmed as comoving with their host star \citep{2014ApJ...781...20K,2014ApJ...780L..30C}.

The star DoAr 21 was also found to be a binary in VLBA measurements by \cite{2008ApJ...675L..29L}. However, the small angular separation of the components (5\,mas) put it outside our region of interest and so it was reobserved and considered a single star in our analysis to avoid biasing our results.

We have included the detection limits from the multiplicity surveys in the literature for both observed multiples and non-detections to provide information on more widely separated companions. Combined, this data covers 80 of our targets.

Two M dwarfs that failed our spectral type cut were observed: EM* SR 22 (M4.5) and SSTc2d J162224.4-245019 (M5). The stars [MMG98] RX J1625.2-2455b, [MMG98] RX J1624.8-2359 and WSB 74 were also observed. These were considered as possible members by several surveys, but have not have their membership firmly established. Finally, Elia 2-29 was observed, despite having no published spectral type. The results of these observations are reported here, but are not included in our statistical analysis or target list.

\subsection{Stellar and companion properties} \label{sec:stellar_properties}

Stellar properties for pre-main sequence stars are difficult to estimate, due to a lack of observational data to constrain stellar evolutionary models. Masses can be uncertain by factors of 2 or more, due to unresolved multiplicity or intrinsic variability. For these reasons, any inferred properties should be treated with caution. Relative properties, such as the mass ratio $q$ are less affected by these systematics and are much less uncertain.

Masses for each target were estimated using the methods outlined in \cite{kraus07a}. Since no single set of stellar models spans the entire range of spectral types found here, it was necessary to combine several independent models. For high mass stars (spectral type F2 and earlier) the temperature scales of \cite{schmidt-kaler82} were used to estimate masses directly from the spectral type. For lower mass stars the effective temperatures were estimated by combining the temperature scales of \cite{schmidt-kaler82} with the M dwarf temperature scales from \cite{luhman03}. The masses were then calculated using the isochrones of \cite{2000ApJ...542..464C} (C00), \cite{baraffe98} (B98) and \cite{siess00} (S00) at 1\,Myr, approximately the median age of Ophiuchus members. The C00 isochrones were used for 0.001-0.1\,M$_\odot$, B98 was used for 0.1-0.5\,M$_\odot$, a mass weighted average of B98 and S00 was used for the range 0.5-1.0\,M$_\odot$, and S00 was used for M $>$ 1.0\,$_\odot$.

Companion masses were calculated using the predicted absolute magnitudes from the isochrones and the measured contrast in the observed band.

To determine which of our targets host disks, we use data from the ``From Molecular Cores to Planet Forming Disks'' Spitzer legacy survey \citep{2003PASP..115..965E}, which covered most of Ophiuchus. One product of this survey was the identification of YSO candidates based on the presence of infrared excesses across SED measurements covering 4.5-24\,$\mu$m\footnote{For futher details on the criteria used to identify disk hosting stars in the Spitzer survey, see \cite{c2dmanual}.}, and we adopt these results.

For those targets without a classification, we use the infrared excess measured by WISE \citep{2010AJ....140.1868W}. We use an approach similar to \cite{2012ApJ...758...31L}, by comparing the $K_s$-$W4$ colour as a function of spectral type for all targets. This metric easily separates the disk hosting population.


\section{Observations and Data Analysis}
\subsection{Observations}
The targets were observed over several years using the Keck NIRC2 instrument with its visible AO system, from 2008 through 2013. One dataset was also taken with the VLT NACO instrument and IR AO system, allowing targets with bright IR magnitudes to be observed. Both instruments have several aperture masks installed in a pupil wheel \citep{2006SPIE.6272E.103T,tuthill2010sparse}. A summary of the observations can be seen in Table \ref{table:obs}.

The observing strategy for our observations was similar to that used in \cite{kraus08}. Targets were observed in groups of $\sim5-19$ objects based on their brightness and location in the sky, with nearby calibrator stars regularly observed to estimate instrument systematics. Calibrators were chosen from the 2MASS catalogue to have similar visible and infrared magnitudes to the targets in each group. In addition, targets that showed no detectable signal were used as calibrators for the remaining targets observed during the same night. In total 63 targets were observed, with total integration times between $8-160$\,s based on the target brightness and chosen to give a similar number of total counts on the detector.

\begin{deluxetable}{ccccp{8.5cm}}
\tabletypesize{\scriptsize}
\tablecaption{Summary of aperture masking observations\label{table:obs}}
\tablehead{\colhead{Date} & \colhead{Telescope} & \colhead{Aperture Mask} & \colhead{Filter} & \colhead{Targets Observed}} 

\startdata
18/06/2008 & Keck & 9 hole & Kp & [MMG98] RX J1620.9-2352, [MMG98] RX J1621.4-2312, WSB 12, GSC 6794-537, 2MASS J16233234-2523485, DoAr 21, EM* SR 3, EM* SR 24 S, EM* SR 21, EM* SR 10, [MMG98] RX J1621.4-2332b, EM* SR 6, [MMG98] RX J1625.2-2455b\\ 
31/05/2009 & Keck & 18 hole & Hcont & HIP 80126 \\
01/06/2009 & Keck & 9 hole & CH4\_short & EM* SR 6, ROXs 47A \\ 
05/04/2010 & Keck & 9 hole & CH4\_short & EM* SR 24 S, ROXs 47A, Haro 1-16\\ 
23/04/2011 & Keck & 9 hole & Kp & DoAr 25, EM* SR 8, WSB 12, DoAr 49, DoAr 24, WSB 46, WLY 2-10, DoAr 32, SSTc2d J162506.9-235050, WSB 63, SSTc2d J163355.6-244205, [MMG98] RX J1624.8-2359, WSB 74\\ 
24/04/2011 & Keck & 9 hole & Kp & ROXs 3, DoAr 25, ROXs 12, GSS 31, WLY 2-10, DoAr 32, DoAr 33, [MMG98] RX J1623.8-2341a, ROXs 45D, SSTc2d J162506.9-235050, WSB 74, [MMG98] RX J1622.7-2325a\\ 
04/06/2011 & Keck & 9 hole & Lp & DoAr 21, EM* SR 21, Haro 1-16\\ 
05/06/2011 & Keck & 9 hole & Lp & DoAr 21, EM* SR 3, EM* SR 21, Haro 1-16\\ 
23/06/2011 & Keck & 9 hole & Kp & EM* SR 24 S, ROXs 42B, EM* SR 6, EM* SR 20\\ 
14/04/2012 & Keck & 9 hole & CH4\_short & EM* SR 24 S, GSS 31, GSS 20, Haro 1-16, [MMG98] RX J1625.3-2402, WSB 40, [WMR2005] 1-21, EM* SR 22, EM* SR 20, GSS 35\\ 
06/05/2012 & Keck & 9 hole & Lp & EM* SR 24 S, DoAr 25, EM* SR 21, GSS 31, DoAr 32, Haro 1-16\\ 
07/07/2012 & Keck & 9 hole & Kp & WLY 2-10, [MMG98] RX J1622.6-2345, [MMG98] RX J1622.8-2333, EM* SR 6\\ 
08/07/2012 & Keck & 9 hole & Kp & DoAr 25, GSS 31, DoAr 33, [MMG98] RX J1623.8-2341a, [MMG98] RX J1625.2-2455b\\ 
09/03/2013 & VLT & 7 hole & Ks & SSTc2d J162224.4-245019, [E2011] 3-37, WLY 1-18, ROXs 39, GSS 29, YLW 47, WSB 52, SSTc2d J162312.5-243641, WLY 1-13, GSS 26, [E2011] 6-62, SSTc2d J163603.9-242344, WSB 9, SSTc2d J162224.4-245019\\
10/03/2013 & VLT & 7 hole & Ks & Elia 2-29, GSS 32, Elia 2-24, [E2011] 1-3, [E2011] 4-29, VSSG 19, [MMG98] RX J1627.2-2404a , [MMG98] RX J1628.2-2405 , [WMR2005] 2-30, [MMG98] RX J1626.3-2407a , [E2011] 3-45, [WMR2005] 2-15, DoAr 13\\
06/08/2013 &  Keck & 9 hole & Kp & ROXs 42B \\
\enddata
\end{deluxetable}

\subsection{Data Analysis and Detection Limits}
The aperture masking data was analysed with a data pipeline developed at the University of Sydney, one of two commonly used software packages for processing such data. A more thorough description of this pipeline can be found in other work \citep[e.g.][]{kraus08}, but a short summary follows. Images are background subtracted, flat fielded and windowed with a super-Gaussian function with a FWHM of 500\,mas, before complex visibilities are measured from the Fourier transform of the image. These raw visibilities are then turned into closure phases using a matched filter approach. Closure phase calibration is achieved by estimating the intrumental closure phases from the average of those measured on the calibrator stars, then subtracting this estimate from the target closure phases.

To determine the limits for companion detection, a Monte-Carlo approach was taken. For each dataset a set of 10,000 random closure phases was drawn from a Gaussian distribution, with a width set by the uncertainty on each closure phase. A model fit was performed for each set of closure phases, yielding a list of 10,000 fake detections that were fit only to noise. To be considered bona-fide, a companion had to have a contrast ratio lower than a 99.9\% (3.3$\sigma$) detection limit, calculated as the contrast ratio for which 99.9\% of the fake detections with a similar separation had a higher contrast.

\begin{figure}
	\begin{center}
	\includegraphics[width=0.49\textwidth]{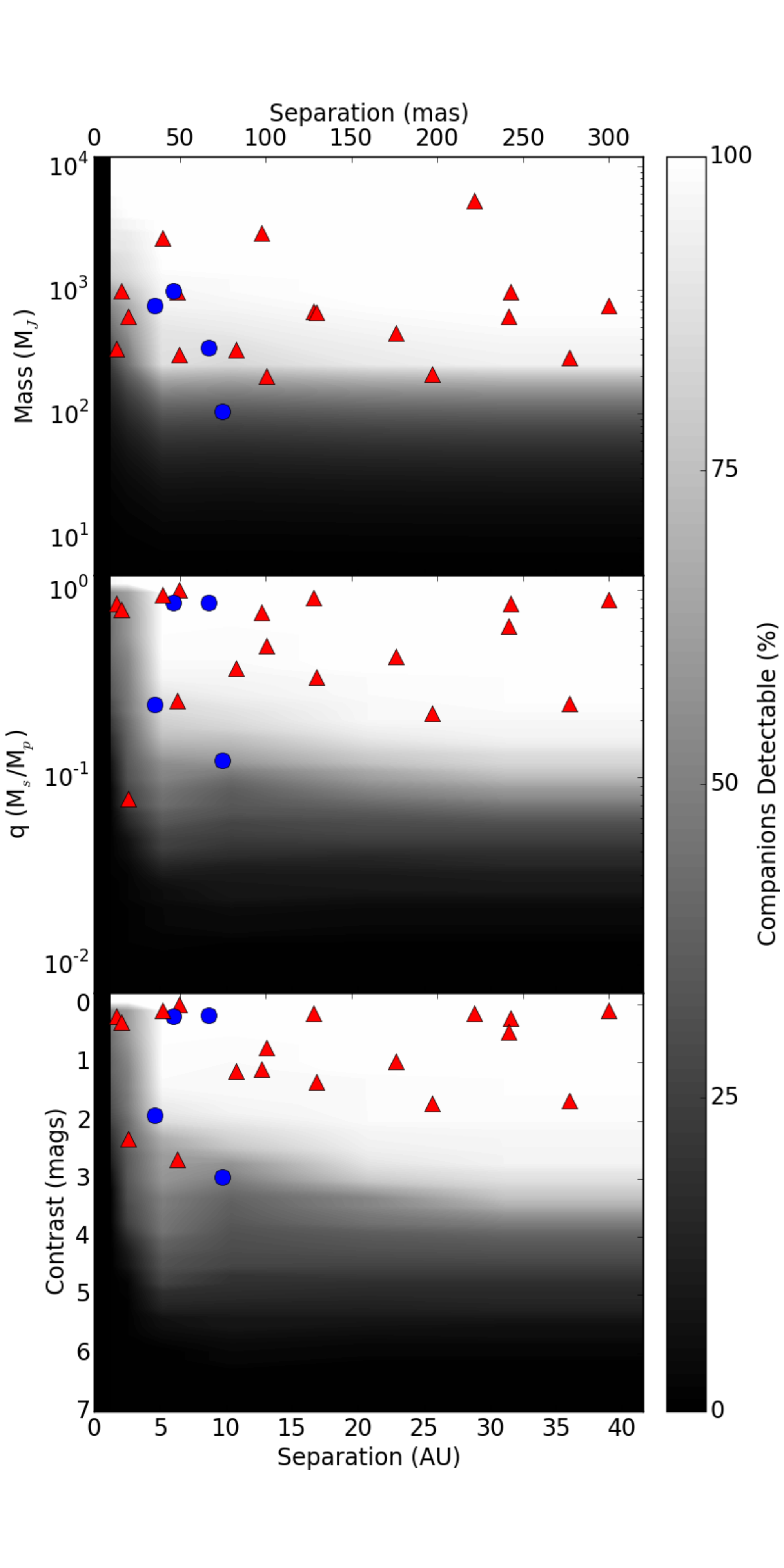}
	\caption{The companion detectability in different regions of the binary parameter space. Masses are calculated using the properties from Table \ref{table:targets} and the approach described in Section \ref{sec:stellar_properties}. Known binaries are marked with red triangles, while the new detections are shown with blue circles. The greyscale levels indicate the number of targets in our sample for which a companion with the given properties would have been detectable.}
	\label{fig:detec_lims}
	\end{center}
\end{figure}

The detection limits calculated from each observation are shown plotted in Figure \ref{fig:detec_lims}, showing the number of targets for which a companion with a given set of parameters could be detected.

For the stars ROXs 47A and [MMG98] RX J1622.7-2325a, a different calibration scheme was used. ROXs 47A star is a known tertiary system from \cite{2003ApJ...591.1064B}. The wider secondary companion was well within the instrument field of view, and appeared to have a significant influence on the measured closure phases of the primary. To account for this, the closure phases from the secondary component were measured and used to calibrate those from the primary. This resulted in a significant improvement in calibration, and a clean detection of the tertiary component. Similarly, [MMG98] RX J1622.7-2325a was a known 0.9\,arcsec binary from \cite{2007ApJ...657..338P}, which placed both components well inside the instrument field of view. Inspection of the images showed that this system was in fact a triple, with a close pair and a brighter primary at a larger separation. Using the measured closure phases from the primary to calibrate those of the close pair produced a clean detection of the tertiary component.

\begin{deluxetable}{lcccccccc}
\tabletypesize{\scriptsize}
\tablecaption{Non Detections\label{tab:non_detecs}}
\tablehead{\colhead{Primary} & \colhead{Date} & \colhead{Filter} & \colhead{} & \colhead{} & \colhead{$\Delta$m} & \colhead{} & \colhead{} & \colhead{} \\ 
\colhead{} & \colhead{} & \colhead{} &  \colhead{10-20} & \colhead{20-40} & \colhead{40-80} & \colhead{80-160} & \colhead{160-240} & \colhead{240-320}} 

\startdata
{DoAr 21} & 18/06/2008 & Kp & 2.08 & 3.78 & 4.65 & 4.51 & 4.32 & 4.35 \\ 
{DoAr 21} & 18/06/2008 & Kp & 1.53 & 3.41 & 4.34 & 4.03 & 4.06 & 4.06 \\ 
{DoAr 21} & 04/06/2011 & Lp & 0.0 & 3.0 & 4.84 & 5.35 & 5.11 & 5.08 \\ 
{DoAr 21} & 05/06/2011 & Lp & 0.0 & 3.18 & 4.98 & 5.49 & 5.3 & 5.26 \\ 
{[MMG98] RX J1620.9-2352} & 18/06/2008 & Kp & 2.15 & 3.82 & 4.69 & 4.58 & 4.38 & 4.4 \\ 
{[MMG98] RX J1620.9-2352} & 18/06/2008 & Kp & 1.53 & 3.41 & 4.34 & 4.05 & 4.06 & 4.06 \\ 
{EM* SR 3} & 18/06/2008 & Kp & 3.01 & 4.53 & 5.36 & 5.28 & 5.22 & 5.28 \\ 
{EM* SR 3} & 05/06/2011 & Lp & 0.0 & 2.81 & 4.67 & 5.17 & 4.98 & 4.95 \\ 
{EM* SR 24 S} & 18/06/2008 & Kp & 0.0 & 0.82 & 2.34 & 1.92 & 1.29 & 0.92 \\ 
{EM* SR 24 S} & 18/06/2008 & Kp & 0.0 & 1.4 & 2.79 & 2.45 & 2.46 & 2.46 \\ 
\enddata
\tablecomments{Angular separation ranges are given in mas.
Table \ref{tab:non_detecs} is published in its entirety in the electronic edition of The Astrophysical Journal. A portion is shown here for guidance regarding its form and content.}
\end{deluxetable}

\begin{deluxetable}{lcccccccc}
\tabletypesize{\scriptsize}
\tablecaption{Companions identified with aperture masking\label{tab:detecs}}
\tablehead{\colhead{Primary} & \colhead{Date} & \colhead{Filter} & \colhead{Separation} & \colhead{P.A.} & \colhead{$\Delta$m} & \colhead{q}  & \colhead{Mass} \\ 
\colhead{} & \colhead{} & \colhead{}  & \colhead{(mas)} & \colhead{(deg)} & \colhead{} & \colhead{$M_s / M_p$} & \colhead{$M_\odot$} } 

\startdata
{[MMG98] RX J1621.4-2332b}\tablenotemark{a} & 18/06/2008 & Kp & 74.98 $\pm$ 0.42 & 50.96 $\pm$ 0.30 & 2.97 $\pm$ 0.03 & 0.12 & 0.10 \\ 
{[MMG98] RX J1626.3-2407a}\tablenotemark{a} & 10/03/2013 & Ks & 66.67 $\pm$ 0.45 & 299.32 $\pm$ 0.45 & 0.19 $\pm$ 0.02 & 0.85 & 0.33 \\ 
{EM* SR 20} & 23/06/2011 & Kp & 50.60 $\pm$ 0.30 & 240.50 $\pm$ 0.30 & 2.68 $\pm$ 0.02 & 0.23 & 0.83 \\ 
{EM* SR 20} & 14/04/2012 & CH4\_short & 48.55 $\pm$ 0.31 & 233.52 $\pm$ 0.36 & 2.66 $\pm$ 0.03 & 0.26 & 0.92 \\ 
{EM* SR 6}\tablenotemark{a} & 18/06/2008 & Kp & 35.60 $\pm$ 0.20 & 270.40 $\pm$ 0.15 & 1.91 $\pm$ 0.01 & 0.24 & 0.72 \\ 
{EM* SR 6}\tablenotemark{a} & 01/06/2009 & CH4\_short & 41.03 $\pm$ 0.22 & 286.46 $\pm$ 0.33 & 2.10 $\pm$ 0.02 & 0.22 & 0.66 \\ 
{EM* SR 6}\tablenotemark{a} & 23/06/2011 & Kp & 57.80 $\pm$ 0.20 & 308.20 $\pm$ 0.22 & 1.94 $\pm$ 0.02 & 0.24 & 0.70 \\ 
{EM* SR 6}\tablenotemark{a} & 07/07/2012 & Kp & 67.00 $\pm$ 1.20 & 314.10 $\pm$ 0.95 & 1.92 $\pm$ 0.07 & 0.24 & 0.71 \\ 
{GSS 35} & 14/04/2012 & Kp & 21.75 $\pm$ 0.39 & 205.11 $\pm$ 0.74 & 1.19 $\pm$ 0.19 & 0.19 & 1.45 \\ 
{GSS 35} & 09/03/2013 & Ks & 28.89 $\pm$ 3.47 & 268.54 $\pm$ 2.13 & 2.26 $\pm$ 0.43 & 0.08 & 0.60 \\ 
{ROXs 39}\tablenotemark{a} & 09/03/2013 & Ks & 46.17 $\pm$ 0.18 & 191.10 $\pm$ 0.80 & 0.20 $\pm$ 0.01 & 0.86 & 0.94 \\ 
{ROXs 47A} & 01/06/2009 & CH4\_short & 51.50 $\pm$ 0.20 & 99.20 $\pm$ 0.30 & 0.19 $\pm$ 0.01 & 0.90 & 2.39 \\ 
{ROXs 47A} & 05/04/2010 & CH4\_short & 43.38 $\pm$ 0.18 & 109.06 $\pm$ 0.19 & 0.22 $\pm$ 0.01 & 0.88 & 2.33 \\ 
{[MMG98] RX J1622.7-2325a}\tablenotemark{a} & 24/04/2011 & Kp & 128.16 $\pm$ 0.28 & 280.53 $\pm$ 0.09 & 0.14 $\pm$ 0.01 & 0.90 & 0.65 \\ 
\enddata
\tablenotetext{a}{Companion reported for the first time in this work.}
\end{deluxetable}


\section{Statistical framework}
To turn the measured detection limits and detections into estimates of the companion frequency distributions, we have employed a statistical framework similar to that of \cite{2006AJ....132.1146C}, \cite{2007ApJ...670.1367L} and \cite{2012ApJ...744..120E}. Further details may be found in these publications and the references therein, but a brief summary follows.

If $f$ is the fraction of stars with a companion in the range of masses $[m_{min},m_{max}]$ and semi-major axes $[a_{min},a_{max}]$, a simple application of Bayes' Theorem results in the following expression for the probability of $f$ given the set of data $\{d_j\}$:
\begin{equation}
P(f|\{d_j\}) = \frac{{\cal L}(\{d_j\} | f) P(f)}{\int^{1}_{0}{\cal L}(\{d_j\} | f) P(f) df}
\end{equation}
Each $d_j$ is equal to 1 if a companion was detected and 0 otherwise. ${\cal L}(\{d_j\} | f)$ is the likelihood of the data. We adopt a prior distribution on $P(f)$ equal to the Jeffreys prior for a Bernoulli trial, $P(f)=\frac{1}{\sqrt{f*\left(1-f\right)}}$, but find no significant difference for any results when adopting a flat prior of $P(f)=1$.

If we let the probability of detecting a companion in the given ranges of masses and semi-major axes, if it was present, be $p_j$, then the probability of detecting a companion around any given star is $fp_j$. Similarly, the probability of not detecting a companion around that star is $1-fp_j$. The calculation of $p_j$ is described in section \ref{sec:pj}.

The likelihood of the data is then given by Equation \ref{eq:likelihood}.
\begin{equation} \label{eq:likelihood}
{\cal L}(\{d_j\} | f) = \prod (1-fp_j)^{1-d_j} (fp_j)^{d_j}
\end{equation}

We then calculate a credible interval for $f$ using the posterior $P(f|\{d_j\})$, such that the interval between $[f_{min},f_{max}]$ contains a fraction $\alpha$ of the total probability. We choose an equal tail distribution, such that the probability contained in $[0,f_{min}]$ and $[f_{max},1]$ are equal. In the case where no companions are found in the stated interval, this is equivalent to solving Equation \ref{eq:upper_bound} for an upper bound on the companion fraction $f_{max}$.

\begin{equation} \label{eq:upper_bound}
\alpha = \int^{f_{max}}_{0} P(f|\{d_j\}) df
\end{equation}

If companions are found, the credible interval is found by solving the following two equations:

\begin{eqnarray}\label{eq:cred_int}
\frac{1-\alpha}{2} = \int^{f_{min}}_{0} P(f|\{d_j\}) df
\\
\frac{1-\alpha}{2} = \int^{1}_{f_{max}} P(f|\{d_j\}) df
\end{eqnarray}

In this work we choose a confidence level of $\alpha=0.68$ for our analysis, equivalent to 1-$\sigma$ limits, unless otherwise stated.

\subsection{Calculation of $p_j$} \label{sec:pj}
In order to calculate the probability of detecting existing companions in the intervals of mass and semi-major axes described above, we perform a Monte Carlo simulation. This involves generating 10,000 companions with masses and angular separations according to known or proposed distributions, and then comparing their properties to the calculated detection limits. The fraction of companions that fall above the mass detection limits is then used as an estimate of $p_j$.

When drawing masses for our simulated companions, we considered two distributions. First is the universal mass function of \cite{2009ApJS..181...62M}, proposed for companions to solar mass stars, given by
\begin{equation}
\frac{dN}{dq} \propto q^{-0.39}
\end{equation}

We also consider a flat mass ratio distribution, consistent with the observed data from \cite{kraus08} and \cite{2011ApJ...731....8K}.

Our results indicate no significant difference in results between these distributions, and only the results of the flat mass ratio distribution are reported here.

Angular separations are calculated by combining information about semi-major axes, eccentricities, orbital phases and inclinations. For the companion eccentricities, we follow the approach of \cite{2012ApJ...744..120E}. In the absence of constraints on the eccentricity distribution of companions, we choose to draw them from a distribution of the form
\begin{equation}
f\left(e\right)=2e
\end{equation}
which is arrived at based on theoretical considerations \citep{ambartsumian1937statistics}.

We convert the semi-major axis $a$ into a projected separation through multiplication by a projection factor $s$, which is calculated from the eccentricity distribution. Combining the above function with the assumption that observing an orbit from any direction is equally likely, \cite{2006ApJ...652.1572B} showed that the cumulative probability distribution for $s$ is well approximated by the function
\begin{equation}
F_s(s)=0.5 \left[1-\cos \left( \frac{\pi}{2}s \right) \right].
\end{equation}
We consider s in the interval $[0,2]$.

Finally, the semi-major axes of our simulated companions are drawn from a distribution of the form
\begin{equation}
\frac{dN}{da} \propto a^{-1}.
\end{equation}
This is consistent with the results of \cite{kraus08,2011ApJ...731....8K}.

Since our approach involves explicitly assuming forms for the distributions of mass and separation, it can introduce systematic biases to the calculated companion fractions. However, since these apply only within each bin of mass and angular separation we can check our assumptions by comparing the observed distributions of these parameters across many bins to those used in our simulations.


\section{Stellar Multiplicity of Ophiuchus Members}

\subsection{Detected Companions}
Our aperture masking observations were sensitive to stellar companions at separations between 10-320\,mas, while the imaging results from the literature were sensitive to binary systems at up to 6\,arcsec. The detection limits for stars that were identified as single stars in our masking observations are listed in Table \ref{tab:non_detecs}.

Eight binary systems were identified from our observations, 5 for the first time, and are listed in Table \ref{tab:detecs}. The known close binary EM* SR 20 was recovered in two epochs, with position angles that suggest at least one complete orbit since its earliest resolved observations in 1990. Both EM* SR 20 and the new multiple system EM* SR 6 have enough epochs to formally allow an orbital fit. In addition, the star [MMG98] RX J1622.7-2325a was shown to be a triple system for the first time.

While observing EM* SR 24 S, it was apparent from the AO guide camera that the contrast ratio between EM* SR 24 S and EM* SR 24 N has changed significantly over time. The S component provided much more flux than expected to the AO wavefront sensor and appeared noticeably brighter than the N component, suggesting that it has changed substantially since the epoch of its UCAC4 photometric measurements. Since this information is not saved during NIRC2 observations, it is not possible to report a quantitative flux measurement. EM* SR 24 S is a known transition disk system, and similar variability is seen in many of these objects.

Our observations failed to recover the close companion to ROXs 42B detected in \cite{1995ApJ...443..625S} and \cite{2005A&A...437..611R} at separations of 56\,mas and 83\,mas respectively. This companion should have been detectable at separations from 10-320\,mas, and by examining the raw images we can extend the outer limit to beyond 1\,arcsec. The non-detection of this companion in our two observation epochs separated by 2 years makes it unlikely that orbital motion may have carried it inwards beyond our inner working angle. We have included the previously reported companion in our analysis, but note that further study is required to determine the nature of this system.

None of the companions reported here have contrasts or calculated masses consistent with brown dwarfs, despite 44 of our 63 masking observations reaching contrasts deep enough to detect such components.

\subsection{Binarity Fraction}

By combining high resolution multiplicity data from the literature with our SAM observations, our survey covers binary and high mass substellar companions separated by 1.3-780\,AU. Overall, we find 36 multiple systems, while 33 members of our target list are single stars across the full range of separations. The remaining 43 targets are missing either imaging or SAM observations, and so have incomplete information. If each observation was able to detect all binaries across this separation range, these numbers suggest an overall binary fraction of $53 \pm 6$\%, ignoring targets that were not observed with at least one of imaging or SAM. Similarly, when restricting the spatial scales to 1.3-41.6\,AU to coincide with those explored by the SAM observations, we find 41 single stars and 22 stars with one or more companions, yielding a binary fraction of $35 \pm 6$\%.

Using the Bayesian techniques described above we are able to include the completeness of coverage of each observed target to better estimate the total companion frequency. Considering a range of separations between 1.3-780\,AU and companion masses greater than 0.08\,M$_\odot$, we find a total companion frequency of 43 $\pm$ 6 \%. This value is between the companion frequency of 35$^{+5}_{-4}$\% found in Upper Scorpius \citep{kraus08} and 64$^{+11}_{-9}$\% in Taurus-Auriga \citep{2011ApJ...731....8K}.

Previous surveys of Ophiuchus members have reported similar companion frequencies. \cite{2005A&A...437..611R} found that $29 \pm 4$\% of targets were in multiple systems over the range 0.13-6.4\,arcsec (17-830\,AU), compared with our value of $32.3 \pm 5.3$\% over the same range. Our results reflect that of \cite{2005A&A...437..611R}, who found significant differences in stellar multiplicity between the Ophiuchus and Taurus-Auriga star forming regions. Despite similarities between the two regions, the companion fraction of Ophiuchus is notably lower.

Across the range of separations explored here, nine of our targets are part of systems with 3 components (EM* SR 13, EM* SR 24 N/S, WSB 18, WSB 38, SSTc2d J162944.3-244122, ROXs 42B, ROXs 43 A/B, ROXs 47A, [MMG98] RX J1622.7-2325a), but we find no evidence of higher order multiplicity across this separation range.

\subsection{The Mass Ratio Distribution}

\begin{figure}[t]
	\begin{center}
	\includegraphics[width=0.49\textwidth]{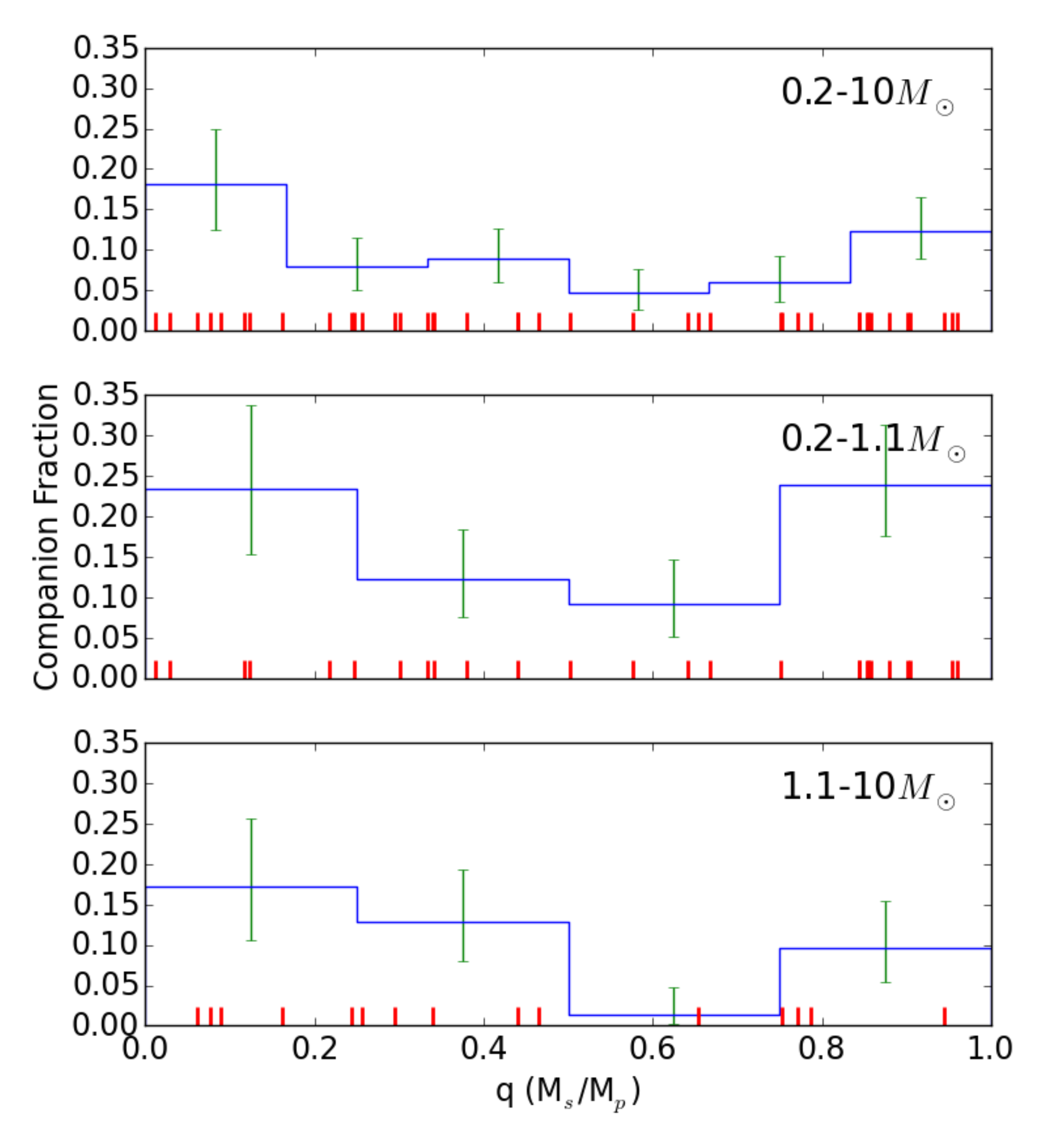}
	\caption{The observed binary fraction of our targets as a function of the mass ratio between the primary and secondary components, for the whole sample (top) as well as the low mass ($<1.1$\,M$_\odot$, middle) and high mass ($>1.1$\,M$_\odot$, bottom) members. Companion separations between 1.3-780\,AU were considered. The individual mass ratios are plotted as vertical lines on the x-axis.}
	\label{fig:f_vs_mass_ratio}
	\end{center}
\end{figure}

A large survey of nearby field dwarfs by \cite{1992ApJ...396..178F} found that the distribution of companion masses is consistent with a flat mass ratio distribution for ratios $q>0.4$. In addition, they found some evidence for a slight excess of equal mass binaries for M types, a conclusion echoed by \cite{1997AJ....113.2246R}. This trend also appears to hold for late M dwarfs and brown dwarfs \citep{2003ApJ...587..407C,2003AJ....126.1526B}. In contrast, the results of \cite{1991A&A...248..485D} indicated a lack of equal mass binaries around F and G type stars. Similarly, \cite{kraus08} found a flat mass ratio distribution holds across the entire range of mass ratios for binary systems in Upper Scorpius, although with no evidence for a peak close to unity.

In Figure \ref{fig:f_vs_mass_ratio} we have shown the observed fraction of stars with companions with separations between 1.3-780\,AU as a function of their mass ratio. The observed distribution appears consistent with a flat mass ratio distribution. We find an excess of equal mass binaries at low significance, echoing the results of \cite{1992ApJ...396..178F} and \cite{1997AJ....113.2246R}. However, we also find a similar excess of low mass ratio companions at low significance.

A one-sided Kolmogorov-Smirnov test shows that the observed distribution is consistent with a flat mass ratio distribution, with test statistic D=0.15 and a p value of 0.21.

Since the companion mass ratio distribution is expected to be dependent on the mass of the primary star, we have shown the results of splitting our target list into a high mass and a low mass subsample. Using a cutoff of 1.1\,$M_\odot$, approximately the median mass of our sample, the resulting mass ratio distributions are also shown in Figure \ref{fig:f_vs_mass_ratio}. We find no significant difference between the shape of the two distributions, although the total companion fraction appears lower for the high mass sample.

\subsection{The Separation Distribution}

Studies of binaries in the field have shown that the separation distribution (or equivalently, the period distribution) is approximately log-normal across a wide range of masses \citep[e.g. ][]{1991A&A...248..485D,1992ApJ...396..178F,2010ApJS..190....1R}. In addition to this, the mean and standard deviation appear to be mass-dependent, with low mass stars having a lower mean separation \citep{1991A&A...248..485D,2003ApJ...587..407C}. However, binaries in young star forming regions may be better matched by a log-flat distribution with an outer cutoff that increases with the mass of the primary \citep{kraus08,2011ApJ...731....8K}.

\begin{figure}[t]
	\begin{center}
	\includegraphics[width=0.49\textwidth]{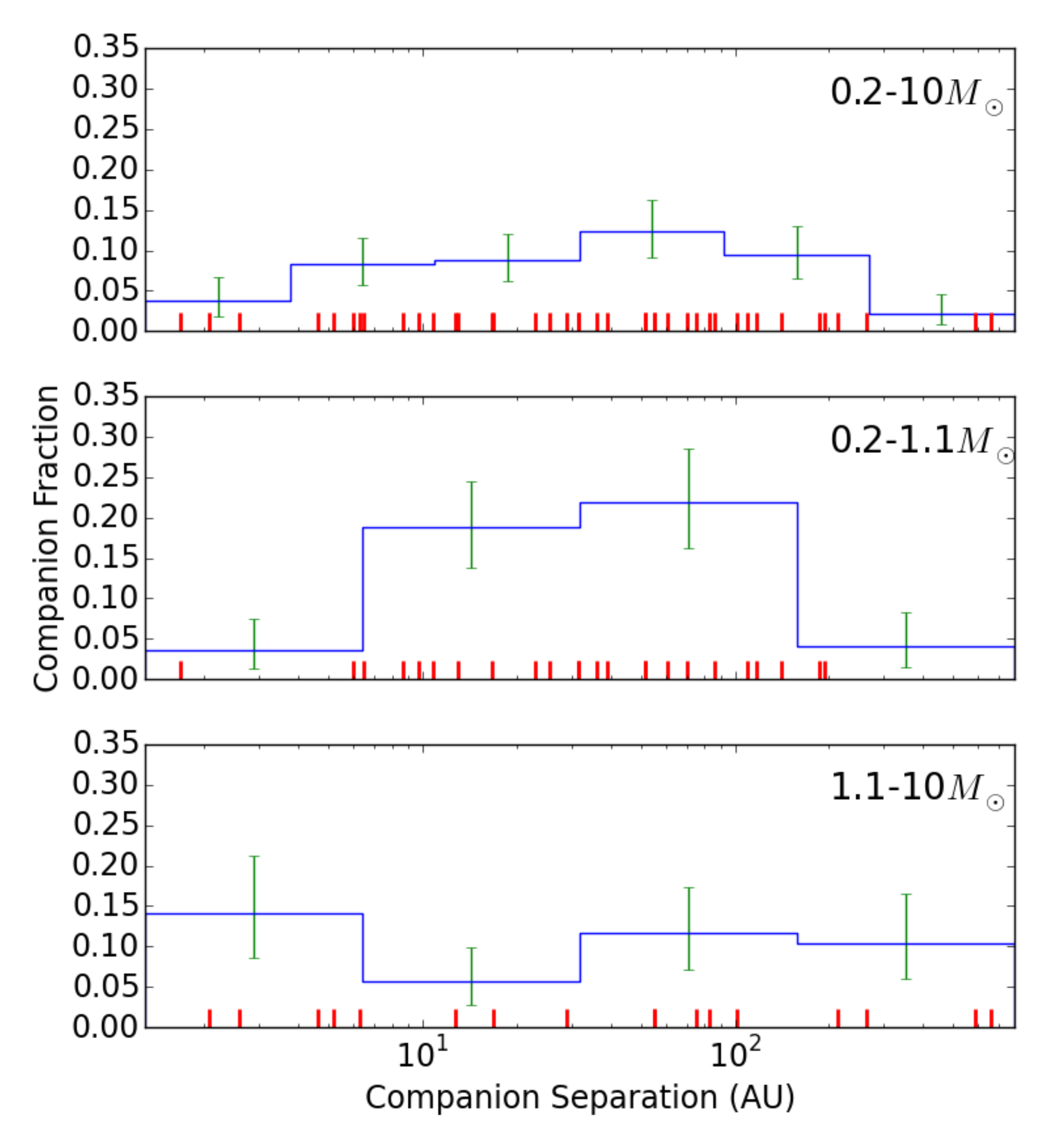}
	\caption{The observed binary fraction of our targets as a function of the separation between the primary and secondary components for the whole sample (top) as well as the low mass ($<1.1$\,M$_\odot$, middle) and high mass ($>1.1$\,M$_\odot$, top) members. Secondary masses between 0.08-6.0 M$_\odot$, and separations between 1.3-780\,AU were considered. The individual separations are plotted as vertical lines on the x-axis.}
	\label{fig:f_vs_sep}
	\end{center}
\end{figure}

Figure \ref{fig:f_vs_sep} shows the observed fraction of stars with companions with masses between 0.08-6\,M$_\odot$ as a function of the separation between the components for our sample. We have opted to plot the minimum observed separation between components (when multiple detection epochs or multiple companions are present), rather than the more physically meaningful semi-major axis of the orbit. For the majority of our binary targets, not enough detection epochs or orbital motion has been observed to allow an orbital solution to be derived, and conversions between projected separations and semi-major axes rely on extrapolated properties of binaries with much smaller separations.

The observed distribution appears log-normal in shape, with a large standard deviation. Over the range of separations explored, there is little evidence to distinguish between the proposed log-normal and log-flat distributions. A one sided Kolmogorov-Smirnov test comparing the observed distribution with a log-flat distribution gives D=0.12 and p=0.46, while the same test on a log-normal distribution gives D=0.11 and p=0.66.

Figure \ref{fig:f_vs_sep} also shows the result of splitting our targets based on primary mass using a cutoff of 1.1\,M$_\odot$, approximately the median mass of the stars in our sample. The separation distribution for the high mass members appears to follow a log-flat separation distribution, consistent with previous results from Upper Scorpius and Taurus \citep{kraus08,2009ApJ...703.1511K,2011ApJ...731....8K}. For the low mass subsample, the distribution appears closer to log-normal in shape, as we find no binary companions with separations between 200-780\,AU and only 1 companion with a separation between 1.3-6\,AU. This difference between the companion distributions of high and low mass members is similar to that found in Taurus by \cite{2011ApJ...731....8K}.

We can use Bayesian analysis techniques to place limits on possible distributions by comparing the observed separations to a log-normal distribution with two parameters: the mean separation $\mu$ and the standard deviation $\sigma$. This approach mirrors that of similar analyses in the literature \citep[e.g.][]{2007ApJ...668..492A,kraus08,2013MNRAS.436.1694R}. The likelihood of the observed separations ${r_i}$ given particular values of $\mu$ and $\sigma$ is given by:
\begin{equation}
L\left( \{r_i\} | \mu, \sigma \right) = \prod_i \frac{\exp \left[ -\left( \log \mu - \log r_i\right)^2 / 2\sigma^2 \right]}
{ \int^{\log r_{\text{max}}}_{\log r_{\text{min}} } \exp \left[ -\left( \log \mu - x \right)^2 / 2\sigma^2 \right]dx}
\end{equation}
where $r_{\text{min}}$ and $r_{\text{max}}$ are the minimum and maximum separations explored, set as 1.3-780\,AU here.

\begin{figure}[h]
	\begin{center} 
	\includegraphics[width=0.49\textwidth]{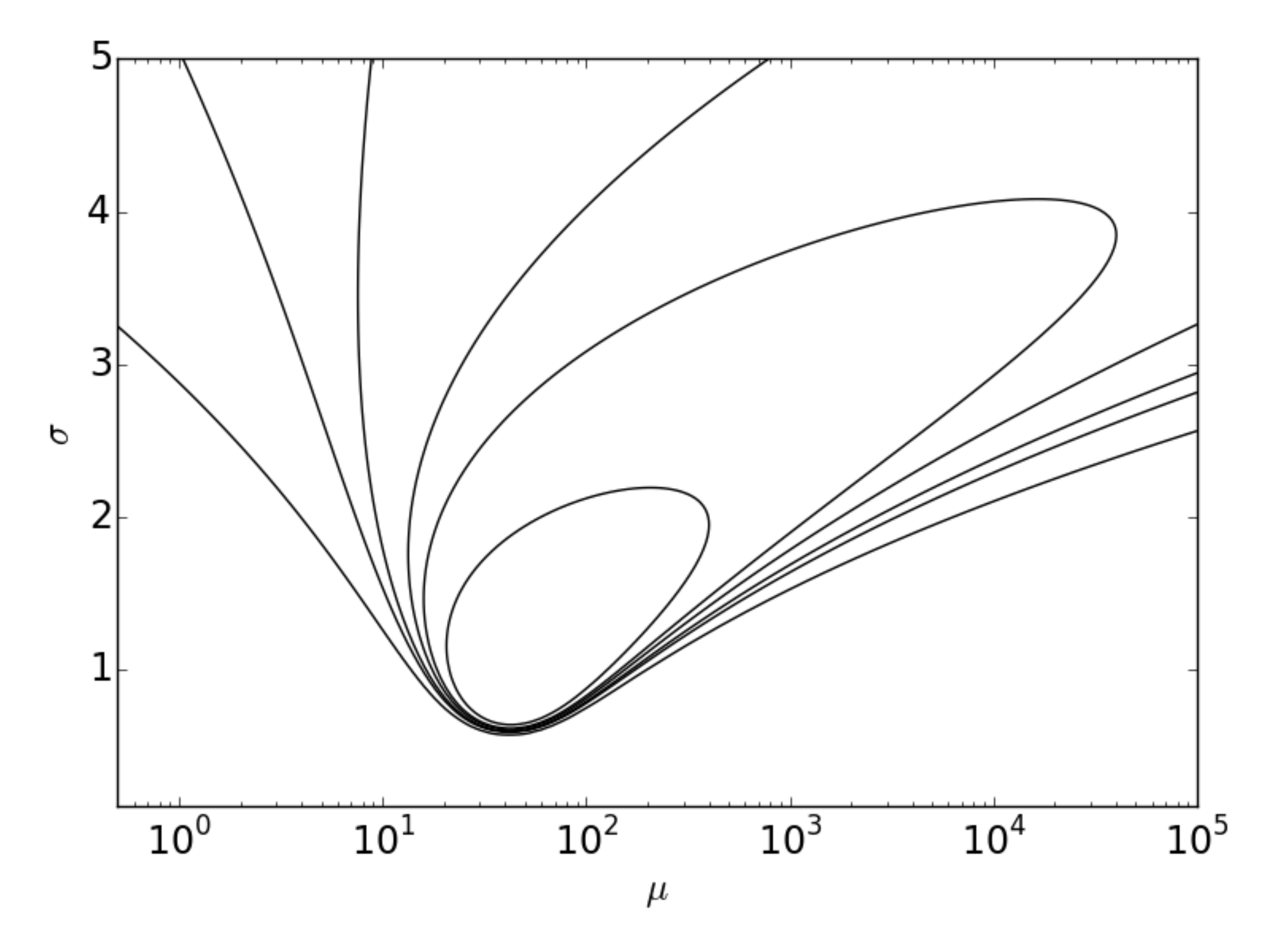}
	\caption{The probability space for the mean separation ($\mu$) and the standard deviation ($\sigma$) assuming a log-normal distribution for the companion separations. Contours enclosing 25\%, 50\%, 75\%, 90\%, 95\% and 99\% of the total probability are shown. The most likely values are $\mu=50$\,AU and $\sigma$=1.0, but a log-flat distribution ($\sigma\gg1$) is not ruled out.}
	\label{fig:mu_sig_prob}
	\end{center}
\end{figure}

Using a uniform prior on $\mu$ and $\sigma$, we find that the most probable values for $\mu$ and $\sigma$ for the full distribution are given by 50 and 1.0 respectively. However, the family of solutions approximating flat distributions ($\sigma \gg 1$, $\mu$ unbound) are within our uncertainties. The probability space for $\mu$ and $\sigma$ is shown in Figure \ref{fig:mu_sig_prob}. Restricting $\sigma$ to be in the range $[0.5,2]$, we find that the mean separation is given by $\mu = 50^{+100}_{-20}$\,AU, while the standard deviation is $\sigma = 1.0^{+0.6}_{-0.2}$. The resulting limits vary substantially based on the choice of prior distribution.

\section{Substellar Companions}
Many of the detection limits of our aperture masking survey reach the brown dwarf mass regime ($\sim$13-80\,\mjup), and combined with the detection limits estimated from the imaging data of previous surveys we can estimate the occurence rate of brown dwarf companions to our targets.

Two of our targets have known widely separated companions that are close to the deuterium burning limit and have been confirmed as co-moving with their host stars. Our mass estimates for these companions places ROXs 42B b in the planetary regime (10\,\mjup) and ROXs 12 b in the brown dwarf regime (25\,\mjup), close to the values of $10\pm4$\,\mjup and $16\pm4$\,\mjup calculated by \cite{2014ApJ...781...20K} (which incorporated the individual estimated ages of these systems).

The companion to the star WSB 28 also has a mass consistent with a high mass brown dwarf (75\,\mjup) and is likely comoving. This companion was identified by \cite{1993A&A...278...81R} and recovered by \cite{2005A&A...437..611R}.

We detect no additional brown dwarf companions to our targets. For companions with masses between 13-80\,\mjup and separations between 1.3-780\,AU, we find an observed frequency of 7$^{+8}_{-5}$\% for our observed targets.

For widely separated companions between 42-780\,AU, we find the frequency to be 4$^{+5}_{-3}$\% while for close companions with separations between 1.3-42\,AU we obtain a 1-$\sigma$ upper limit of 12\%.


\section{The Effects of Multiplicity on Disk Evolution}

Combining our census on the binarity of our targets with information on which stars host disks provides a fantastic opportunity to study the ways in which multiplicity affects the evolution of disks around young stars.

Since 17 of our targets were selected based on the infrared surveys of \cite{luhman99} and \cite{cieza10}, their inclusion may introduce a bias towards disk hosting stars with infrared excesses and so we have not included them in the analysis presented in this section.

We find that 44 of our targets host disks, while 52 have no disk, and 1 has insufficient information to classify it. This gives an overall disk frequency of $46 \pm 5$\%, including those stars with no multiplicity information. However, this figure changes substantially depending on the binarity of the target. For binary targets, the ratio is $50\pm8$\%, while for confirmed single stars it is $70\pm9$\%.

\begin{figure*}
	\begin{center}
	\includegraphics[width=0.9\textwidth]{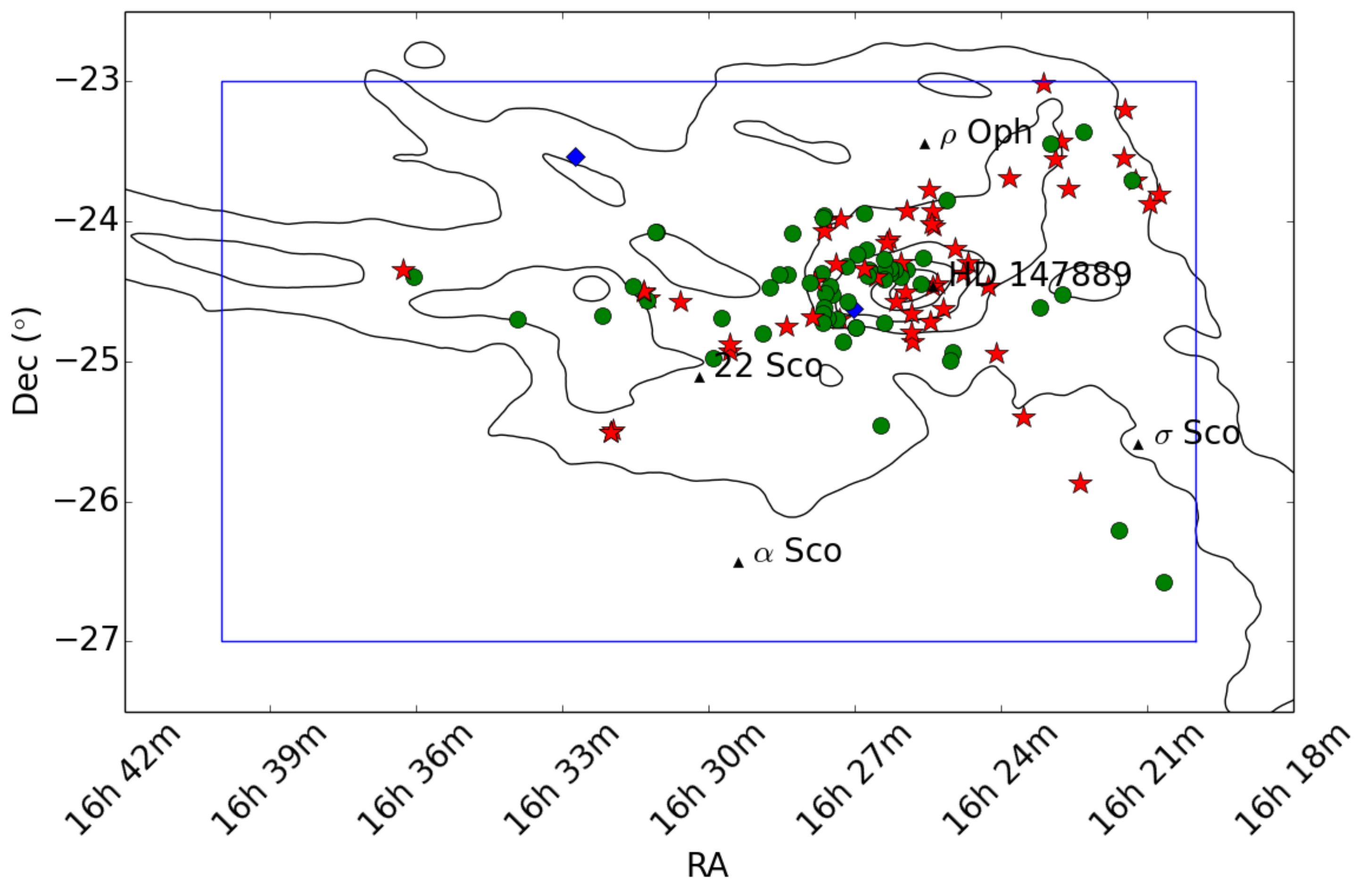}
	\caption{The spatial distribution of stars in our survey, labelled according to whether they possess circumstellar disks. Targets with no disk are marked with red stars, those with disks are labelled with green circles and targets with insufficient information to classify them are marked with blue diamonds. A high fraction of stars in the highest density regions have disks, while the disk fraction is lower in the West and North-West. Stars in these regions are expected to be older than those in the core, and many have lost their disks.}
	\label{fig:dust_map_with_disks}
	\end{center}
\end{figure*}

Stars with incomplete multiplicity information have a disk frequency of $24 \pm 7$\%. This low disk frequency is likely due to the spatial distribution of those targets. Previous multiplicity surveys have targeted the central region of the L1688 cloud, which hosts the youngest stars in our sample. Evidence suggests that star formation has been ongoing in some regions of the Ophiuchus cloud complex for as long as 10\,Myr \citep{martin98}, and the targets furthest from the L1688 central region show a correspondingly lower disk fraction. The spatial distribution of our targets and their disk hosting status are shown in Figure \ref{fig:dust_map_with_disks}.

\begin{figure}[ht]
	\begin{center}
	\includegraphics[width=0.49\textwidth]{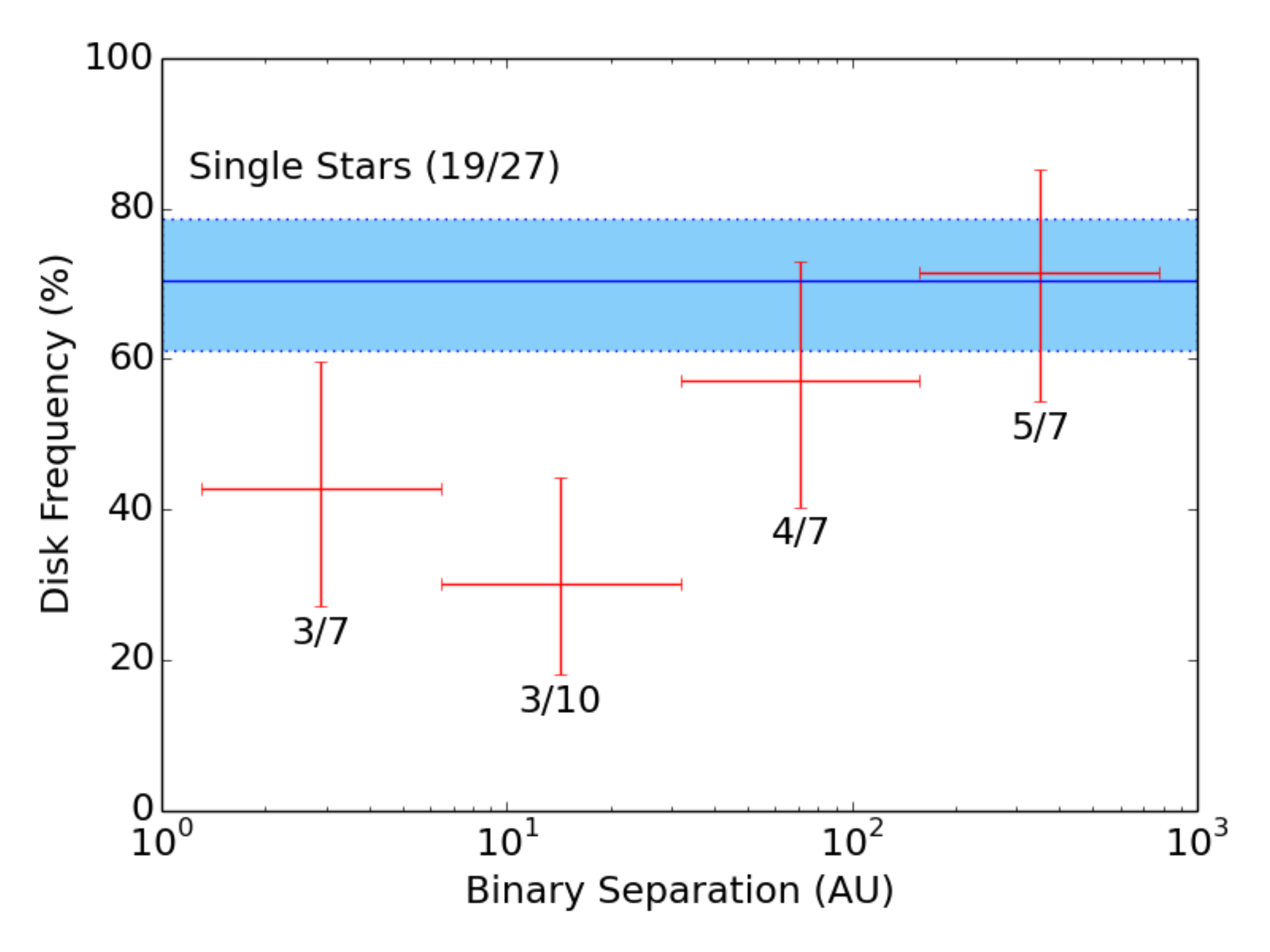}
	\caption{The fraction of observed binary stars that host disks as a function of the separation between the host star and the binary companion. }
	\label{fig:disk_freq_vs_binary_sep}
	\end{center}
\end{figure}

Figure \ref{fig:disk_freq_vs_binary_sep} shows the result of splitting the stars based on the separation of the binary companions. From our survey we find that close binary systems ($<$40\,AU) are less likely to host disks than wider binaries and single stars. This implies that the presence of a close binary companion either speeds up the process of disk dispersal or inhibits its formation. This conclusion echoes that of \cite{2012ApJ...745...19K}, who found that only $\sim1/3$ of stars with companions at separations smaller than 40\,AU hosted disks at ages of only 1-2\,Myr. At the 1-2\,Myr age of Ophiuchus, we find a similar result.

At the age of our sample, stars with a wide companion ($>$40\,AU) have a similar disk frequency as single stars ($69\pm12$\% compared to $70\pm9$\%). This implies that the presence of a wide companion does not signicantly affect disk evolution at an age of $\sim$1-2\,Myr.

The fraction of stars that host circumstellar disks at such young ages is of critical importance for the formation of giant planets. The core accretion mode of star formation \citep{1996Icar..124...62P} requires several Myr to efficiently produce giant planets. Since only $\sim1/3$ of close binary systems in star forming regions still possess disks at an age of only 1-2\,Myr, this implies that giant planets around such systems would be rare. This supports early results from radial velocity and transit surveys \citep{2007A&A...462..345D,2012A&A...542A..92R,2014ApJ...791..111W}.

A significant population of our targets have SED gaps indicative of a transition disk. 13 transition disk candidates were identified by \cite{cieza10}: WSB 12, SSTc2d J162506.9-235050, DoAr 25, DoAr 32, DoAr 33, WSB 63, SSTc2d J163355.6-244205, SSTc2d J162245.4-243124, WSB 38, EM* SR 9, SSTc2d J162944.3-244122, ROXs 42C and DoAr 21. In addition, 4 more stars have been identified as transition disks: EM* SR 24 S, Haro 1-16 (DoAr 44), EM* SR 21 and YLW 46 (Oph IRS 48) \citep{2011ApJ...732...42A,2007A&A...469L..35G}.

Of these 17 targets, 4 have close companions ($\lesssim40$\,AU) and may be circumbinary rather than transitional (DoAr 21, ROXs 42C, SSTc2d J162944.3-244122, WSB 38). SSTc2d J162944.3-244122 and WSB 38 are known triple systems with a close binary and wider tertiary, and it is unclear which component hosts the disk. In addition, EM* SR 24 S, SSTc2d J162245.4-243124, EM* SR 9 and EM* SR 21 have companions at wider separations that may not be the source of their disk gaps. However, the remaining 9 targets have no detected stellar or substellar companion.

Many scenarios have been proposed for the origin of the disk gaps in transition disks. The favoured scenario invokes dust clearing due to the presence of giant planets forming in the disk. However, for the majority of the transition disk candidates in our survey we find a lack of stellar or brown dwarf companions capable of causing such gaps. This lends weight to the giant planet formation hypothesis, showing that most of these disks are not circumbinary in nature.


\section{Conclusion}

We have performed a multiplicity survey of the $\rho$ Ophiuchus cloud complex, studying the occurrence rates of stellar and sub-stellar companions. This information was combined with knowledge of circumstellar disks among our targets from Spitzer to investigate how these properties relate. Our results point to several significant conclusions for the processes of star and giant planet formation.

\begin{enumerate}
\item We have identified 36 multiple systems, including 9 systems with 3 components. Of these, 5 stellar companions were resolved for the first time. This gives a binarity fraction of 43$\pm6$\%, intermediate between the results of similar star forming regions in Upper Scorpius \citep{kraus08} and Tauris-Auriga \citep{2011ApJ...731....8K}.

\item The observed distribution of companion masses is consistent with a flat mass ratio distribution when considering components with separations between 1.3-780\,AU, while the separation distribution is consistent with a log-normal or log-flat distribution over the same range when considering companions with masses between 0.08-6.0\,M$_\odot$. These results agree with previous surveys of star forming regions and the field.

\item The fraction of disk hosting stars depends strongly on the existence of a close ($\lesssim$40\,AU) stellar-mass companion. Only $\sim$1/3 of stars in close binary systems still host circumstellar disks after only 1-2\,Myr, suggesting that the presence of such companions significantly speeds up the process of disk dispersal or inhibits their formation. The lack of disks around close binary systems suggests that giant planets formed by the slow process of core accretion should be rare around such systems, a prediction supported by the results of early RV and transit surveys.

\item In contrast, a more widely separated companion appears to have no effect on the presence of a disk at 1-2\,Myr. We find that $\sim70$\% of both single stars and wide binaries retain their circumstellar disks at the age of Ophiuchus members.

\end{enumerate}

\acknowledgements

The authors wish to recognize and acknowledge the very significant cultural role and reverence that the summit of Mauna Kea has always had within the indigenous Hawaiian community. We are most fortunate to have the opportunity to conduct observations from this mountain.

L.C. was supported by ALMA-CONICYT \#31120009, CONICYT-FONDECYT \#1140109, and the Millennium Science Initiative (Chilean Ministry of Economy), through grant Nucleus RC130007.

{\footnotesize
\bibliographystyle{apj}
\bibliography{bdd_bib,trans_disks}
}

\end{document}